\documentclass[aps,prl,twocolumn,showpacs,showkeys,amsmath,amssymb]{revtex4}
\usepackage{amsfonts}
\usepackage{amsmath}
\usepackage{graphicx}
\usepackage{subfigure}
\usepackage{dcolumn}
\usepackage{bm}
\usepackage{booktabs}
\usepackage{graphicx,graphics,dcolumn,booktabs,bm}
\usepackage{longtable,lscape}
\pdfoutput=1
\usepackage{txfonts}
\usepackage{overpic}
\usepackage{amssymb}
\usepackage{indentfirst}
\usepackage{epsfig}
\usepackage{feynmf}   
\usepackage{epstopdf}   
\usepackage{slashed}  
\usepackage{multirow}

\usepackage{ulem}
\usepackage[usenames,dvipsnames]{color}

\begin{document}

\title{Towards exotic hidden-charm pentaquarks in QCD}

\author{Hua-Xing Chen$^1$}
\author{Wei Chen$^2$}
\email{wec053@mail.usask.ca}
\author{Xiang Liu$^{3,4}$}
\email{xiangliu@lzu.edu.cn}
\author{T. G. Steele$^2$}
\email{tom.steele@usask.ca}
\author{Shi-Lin Zhu$^{5,6,7}$}
\email{zhusl@pku.edu.cn}
\affiliation{
$^1$School of Physics and Nuclear Energy Engineering and International Research Center for Nuclei and Particles in the Cosmos, Beihang University, Beijing 100191, China
\\
$^2$Department of Physics and Engineering Physics, University of Saskatchewan, Saskatoon, SK, S7N 5E2, Canada
\\
$^3$School of Physical Science and Technology, Lanzhou University, Lanzhou 730000, China
\\
$^4$Research Center for Hadron and CSR Physics, Lanzhou University and Institute of Modern Physics of CAS, Lanzhou 730000, China
\\
$^5$School of Physics and State Key Laboratory of Nuclear Physics and Technology, Peking University, Beijing 100871, China
\\
$^6$Collaborative Innovation Center of Quantum Matter, Beijing 100871, China
\\
$^7$Center of High Energy Physics, Peking University, Beijing 100871, China
}

\begin{abstract}
Inspired by $P_c(4380)$ and $P_c(4450)$ recently observed by LHCb, a QCD sum rule investigation is performed, by which they can be identified as exotic hidden-charm pentaquarks composed of an anti-charmed meson and a charmed baryon. Our results suggest that $P_c(4380)$ and $P_c(4450)$ have quantum numbers $J^P=3/2^-$ and $5/2^+$, respectively. Furthermore, two extra hidden-charm pentaqurks with configurations $\bar{D}\Sigma_c^*$ and $\bar{D}^*\Sigma_c^*$ are predicted, which have spin-parity quantum numbers $J^P=3/2^-$ and $J^P=5/2^+$, respectively.
As an important extension, the mass predictions of hidden-bottom pentaquarks are also given. Searches for these partners of $P_c(4380)$ and $P_c(4450)$ are especially accessible at future experiments like LHCb and BelleII.
\end{abstract}

\maketitle

{\it Introduction}.---Exploring exotic matter beyond conventional hadron configurations is one of the most intriguing current research topics of hadronic physics, and these studies will improve our understanding of non-perturbative QCD. With the experimental progress on this issue over the past decade, dozens of $XYZ$ charmonium-like states have been reported, which provide us good opportunities to identify exotic hidden-charm four-quark matter \cite{Liu:2013waa}. Facing such abundant novel phenomena relevant to four-quark matter, we naturally conjecture that there should exist hidden-charm pentaquark states \cite{Wu:2010jy,Yang:2011wz,Li:2014gra,Uchino:2015uha,Karliner:2015ina}. In fact, the possible hidden-charm molecular pentaquarks composed of an anti-charmed meson and a charmed baryon were investigated systematically within the one boson exchange model in Ref. \cite{Yang:2011wz}. However, the experimental evidence of exotic hidden-charm pentaquark state has been absent until the LHCb Collaboration's recent observations of two hidden-charm pentaquark resonances.

Via the $\Lambda_b\to J/\psi p K$ process, LHCb observed two
enhancements, $P_c(4380)$ and $P_c(4450)$, in the $J/\psi p$ invariant mass spectrum \cite{lhcb}, which shows that they must have hidden-charm quantum number and isospin $I=1/2$. Additionally, their resonance parameters are measured, i.e., $M_{P_c(4380)}=4380\pm 8\pm 29\, \mathrm{MeV}$, $\Gamma_{P_c(4380)}=205\pm18\pm86\, \mathrm{MeV}$,
$M_{P_c(4450)}=4449.8\pm 1.7\pm 2.5 \,\mathrm{MeV}$, and $\Gamma_{P_c(4450)}=39\pm5\pm19\, \mathrm{MeV}$ \cite{lhcb}.
Later, they are studied by using the boson exchange model~\cite{Chen:2015loa} and the topological soliton model~\cite{Scoccola:2015nia}, etc.


In this letter, we give an explicit QCD sum rule investigation to $P_c(4380)$ and $P_c(4450)$.
We shall investigate the possibility of interpreting them as hidden-charm pentaquark configurations composed of an anti-charmed meson and a charmed baryon: $P_c(4380)$
can be well reproduced using a $[\bar D^* \Sigma_c]$ structure with quantum numbers $J^P=3/2^-$, and $P_c(4450)$ can be well reproduced using a mixed structure
of $[\bar D^* \Lambda_c]$ and $[\bar D \Sigma_c^*]$ with $J^P=5/2^+$. One notes that the ``structure'' here means we are using 
meson-baryon currents having the color configuration $[\bar c_d q_d][\epsilon^{abc}c_a q_b q_c]$, where $a \cdots d$
are color indices, $q$ represents $up$, $down$ and $strange$ quarks, and $c$ represents a $charm$ quark. These local currents could probe either a tightly-bound pentaquark structure
or a molecular structure composed of an anti-charmed meson and a charmed baryon.

Besides clarifying properties of these two observed $P_c(4380)$ and $P_c(4450)$ pentaquarks, in this letter we further give theoretical predictions of two extra hidden-charm pentaqurks with configurations $\bar{D}\Sigma_c^*$ and $\bar{D}^*\Sigma_c^*$, as partners of $P_c(4380)$ and $P_c(4450)$. After the LHCb's observation  \cite{lhcb}, experimental exploration to these predicted hidden-charm pentaquarks will be an intriguing research topic, of interest  to both experimentalists and theorists.

{\it Interpretation of} observed $P_c(4380)$ and $P_c(4450)$ states.---As the first step, we briefly discuss how to construct local pentaquark interpolating currents having spin $J = 3/2$, flavor-octet $\mathbf{8}_F$, and containing one $c \bar c$ pair. There are two possible color configurations, either $[\bar c_d c_d][\epsilon^{abc}q_a q_b q_c]$ or $[\bar c_d q_d][\epsilon^{abc}c_a q_b q_c]$.
These two configurations, if they are local, can be related by the Fierz transformation as well as the color rearrangement:
\begin{eqnarray}
\delta^{de} \epsilon^{abc} &=& \delta^{da} \epsilon^{ebc} + \delta^{db} \epsilon^{aec} + \delta^{dc} \epsilon^{abe} \, .
\label{eq:cr}
\end{eqnarray}
The former configuration, $[\bar c_d c_d][\epsilon^{abc}q_a q_b q_c]$, can be easily constructed based on the results of Ref.~\cite{Chen:2008qv} that there are three independent local light baryon fields of flavor-octet and having a positive parity:
\begin{eqnarray}
\nonumber N^N_1 &=& \epsilon_{abc} \epsilon^{ABD} \lambda_{DC}^N (q_A^{aT} C q_B^b) \gamma_5 q_C^c \, ,
\\ N^N_2 &=& \epsilon_{abc} \epsilon^{ABD} \lambda_{DC}^N (q_A^{aT} C \gamma_5 q_B^b) q_C^c \, ,
\label{eq:baryon}
\\ \nonumber N^N_{3\mu} &=& \epsilon_{abc} \epsilon^{ABD} \lambda_{DC}^N (q_A^{aT} C \gamma_\mu \gamma_5 q_B^b) \gamma_5 q_C^c \, ,
\end{eqnarray}
where $A \cdots D$ are flavor indices, and $q_{A}=(u\, ,d\, ,s)$ is the light quark field of flavor-triplet. Together with light baryon fields having negative parity, $\gamma_5 N^N_{1,2}$ and $\gamma_5 N^N_{3\mu}$, and the charmonium fields:
\begin{eqnarray}
\nonumber &\bar c_d c_d \, [0^+] \, , \bar c_d \gamma_5 c_d \, [0^-] \, ,&
\\ \nonumber &\bar c_d \gamma_\mu c_d \, [1^-] \, , \bar c_d \gamma_\mu \gamma_5 c_d \, [1^+] \, , \bar c_d \sigma_{\mu\nu} c_d \, [1^\pm] \, ,&
\end{eqnarray}
we can construct the currents containing $J=3/2$ components, which are:
\begin{eqnarray}
\nonumber & [\bar c_d c_d] [ N^N_{3\mu}] \, , [\bar c_d \gamma_5 c_d] [ N^N_{3\mu}] \, ,  [\bar c_d \gamma_\mu c_d] [ N^N_{1,2}] \, , &
\\ & [\bar c_d \gamma_\mu \gamma_5 c_d] [ N^N_{1,2}] \, , [\bar c_d \gamma_\mu c_d] [ N^N_{3\nu}] \, , [\bar c_d \gamma_\mu \gamma_5 c_d] [ N^N_{3\nu}] \, , &
\label{currents}
\\ \nonumber & [\bar c_d \sigma_{\mu\nu} c_d] [ N^N_{1,2}] \, , [\bar c_d \sigma_{\mu\nu} c_d] [ N^N_{3\rho}] \, , &
\end{eqnarray}
as well as their partners having opposite parities, i.e., $[\cdots] [\gamma_5 \cdots]$ (such as $[\bar c_d c_d] [\gamma_5 N^N_{3\mu}]$). We note that their parities
are a bit complicated and will be discussed later.

Besides $J=3/2$ components, these currents can also contain $J=1/2$ and $5/2$ components. The $J=1/2$ components can be safely removed in the two-point correlation functions, which will be discussed after Eq.~(\ref{piDsSs52}) and so we shall not consider any more; to separate $J=3/2$ and $5/2$ components, we need to use projection operators. For example, the current
\begin{eqnarray}
\eta^N_{3\mu\nu} &=& [\bar c_d \gamma^\mu c_d] [ \gamma_5 N^N_{3\nu}] \, ,
\end{eqnarray}
contains both spin $J = 3/2$ and $5/2$ components:
\begin{eqnarray}
\nonumber \eta_{3[\mu\nu]}^{N} &=& [\bar c_d \gamma^\mu c_d] [ \gamma_5 N^N_{3\nu}] - [\bar c_d \gamma^\nu c_d] [ \gamma_5 N^N_{3\mu}] \, ,
\\ \eta_{3\{\mu\nu\}}^{N} &=& [\bar c_d \gamma^\mu c_d] [ \gamma_5 N^N_{3\nu}] + [\bar c_d \gamma^\nu c_d] [ \gamma_5 N^N_{3\mu}]\, ,
\label{eq:current3}
\end{eqnarray}
where $\eta_{3[\mu\nu]}^{N}$ contains both $J^P={3/2}^+$ and ${3/2}^-$ components, and $\eta_{3\{\mu\nu\}}^{N}$ contains only the $J^P={5/2}^+$ component.

Among the currents listed in Eqs.~(\ref{currents}) and (\ref{eq:current3}), $\eta^N_{1,2\mu} \equiv [\bar c_d \gamma_\mu c_d] [ N^N_{1,2}]$ of $J^P = {3/2}^-$ couples well to the combination of $J/\psi$ and proton through $S$-wave, and
$\eta_{3\{\mu\nu\}}^{N}$ of $J^P = {5/2}^+$ couples well to the combination of $J/\psi$ and $proton$ through $P$-wave, when their quark contents are $c \bar c uud$:
\begin{eqnarray}
\nonumber \eta_{1\mu}^{c \bar c uud} &=& [\bar c_d \gamma_\mu c_d]  [\epsilon_{abc} (u^T_a C d_b) \gamma_5 u_c] \, ,
\\ \nonumber \eta_{2\mu}^{c \bar c uud} &=& [\bar c_d \gamma_\mu c_d]  [\epsilon_{abc} (u^T_a C \gamma_5 d_b) u_c] \, ,
\\ \eta_{3\{\mu\nu\}}^{c \bar c uud} &=& [\bar c_d \gamma_\mu c_d] [\epsilon_{abc} (u^T_a C \gamma_\nu \gamma_5 d_b) u_c] + \{ \mu \leftrightarrow \nu \} \, .
\end{eqnarray}
In the following we shall use the mixed current containing ``Ioffe's baryon current'', which couples strongly to the lowest-lying nucleon state~\cite{Belyaev:1982sa,Espriu:1983hu}
\begin{eqnarray}
\eta_{12\mu}^{c \bar c uud} &=& \eta_{1\mu}^{c \bar c uud} - \eta_{2\mu}^{c \bar c uud} \, ,
\end{eqnarray}
as well as $\eta_{3\{\mu\nu\}}^{c \bar c uud}$ to perform QCD sum rule analyses. However, we shall see that the results are not useful.

Considering that the experimental observed states have masses significantly larger than the threshold of $J/\psi$ and proton, but close to thresholds of $D/D^*$ and $\Lambda_c/\Sigma_c/\Sigma_c^*$, we shall also construct  currents belonging to the other configuration, $[\bar c_d q_d][\epsilon^{abc}c_a q_b q_c]$, and use them to perform QCD sum rule analyses. Because currents of this type can not be systematically constructed so easily, we just choose some of them and give their relations to $\eta_{1,2\mu}^{c \bar c uud}$ and $\eta_{3\{\mu\nu\}}^{c \bar c uud}$, but leave the detailed discussions for our future studies.

We can transform the current $\eta_{12\mu}^{c \bar c uud}$ using the Fierz transformation (f.t.) and the color rearrangement (c.r.) to be
\begin{eqnarray}
\eta_{12\mu}^{c \bar c uud} &\xrightarrow{f.t.\&c.r.}& {1\over8} J^{\bar D^*\Sigma_c}_{\mu} + {1\over8} J^{\bar D\Sigma_c^*}_{\mu} + \cdots \, ,
\label{eq:relation12}
\end{eqnarray}
where
\begin{eqnarray}
J^{\bar D^*\Sigma_c}_{\mu} &=& [\bar c_d \gamma_\mu d_d]  [\epsilon_{abc} (u^T_a C \gamma_\nu u_b) \gamma^\nu \gamma_5 c_c]  \, ,
\label{DsS32}
\\
J^{\bar D\Sigma_c^*}_{\mu} &=& [\bar c_d \gamma_5 d_d]  [\epsilon_{abc} (u^T_a C \gamma_\mu u_b) c_c]  \, .
\label{DSs32}
\end{eqnarray}
The former one, $J^{\bar D^*\Sigma_c}_{\mu}$, seems to contain color-singlet $\bar D^*$ and $\Sigma_c$, which structure we denote as $[\bar D^*\Sigma_c]$. It may be interpreted as a tightly-bound pentaquark structure or a $[\bar D^*\Sigma_c]$ molecular state. If there exists a state with such structures, this current would couple strongly to it. The latter one, $J^{\bar D\Sigma_c^*}_{\mu}$, has a $[\bar D\Sigma_c^*]$ structure.

We can also transform the current $\eta_{3\{\mu\nu\}}^{c \bar c uud}$ to be
\begin{eqnarray}
\eta_{3\{\mu\nu\}}^{c \bar c uud} &\xrightarrow{f.t.\&c.r.}& -{1\over8} J^{\bar D^*\Sigma_c^*}_{\{\mu\nu\}} -{1\over8} J^{\bar D\Sigma_c^*}_{\{\mu\nu\}} -{3\over8} J^{\bar D^*\Lambda_c}_{\{\mu\nu\}} + \cdots \, ,
\label{eq:relation3}
\end{eqnarray}
where
\begin{eqnarray}
J^{\bar D^*\Sigma_c^*}_{\{\mu\nu\}} &=& [\bar c_d \gamma_\mu d_d] [\epsilon_{abc} (u^T_a C \gamma_\nu u_b) \gamma_5 c_c] + \{ \mu \leftrightarrow \nu \} \, ,
\label{DsSs52}
\\ J^{\bar D\Sigma_c^*}_{\{\mu\nu\}} &=& [\bar c_d \gamma_\mu \gamma_5 d_d] [\epsilon_{abc} (u^T_a C \gamma_\nu u_b) c_c] + \{ \mu \leftrightarrow \nu \} \, ,
\label{DSs52}
\\ J^{\bar D^*\Lambda_c}_{\{\mu\nu\}} &=& [\bar c_d \gamma_\mu u_d] [\epsilon_{abc} (u^T_a C \gamma_\nu \gamma_5 d_b) c_c] + \{ \mu \leftrightarrow \nu \} \, .
\label{DsL52}
\end{eqnarray}
They have $\bar D^*\Sigma_c^*$, $\bar D\Sigma_c^*$ and $\bar D^*\Lambda_c$ structures, respectively.

In the following, we shall use the method of QCD sum rules~\cite{Shifman:1978bx,Reinders:1984sr,Nielsen:2009uh,Chen:2015ata} to investigate $\eta_{12\mu}^{c \bar c uud}$ and $\eta_{3\{\mu\nu\}}^{c \bar c uud}$
for the $[\bar c_d c_d][\epsilon^{abc}q_a q_b q_c]$ structure, and $J^{\bar D^*\Sigma_c}_{\mu}$, $J^{\bar D\Sigma_c^*}_{\mu}$, $J^{\bar D^*\Sigma_c^*}_{\{\mu\nu\}}$, $J^{\bar D\Sigma_c^*}_{\{\mu\nu\}}$, and $J^{\bar D^*\Lambda_c}_{\{\mu\nu\}}$ for the $[\bar c_d q_d][\epsilon^{abc}c_a q_b q_c]$ structure. Eqs.~(\ref{eq:relation12}) and (\ref{eq:relation3}) suggest that the structures coupled by these currents, if exist, would naturally decay to $J/\psi$ and $proton$ final states:
$J^{\bar D^*\Sigma_c}_{\mu}$ and $J^{\bar D\Sigma_c^*}_{\mu}$ couple equally to ``$S$-wave'' $J/\psi$ and $proton$, and  $J^{\bar D^*\Lambda_c}_{\{\mu\nu\}}$ couples to ``$P$-wave'' $J/\psi$ and $proton$ more strongly than $J^{\bar D^*\Sigma_c^*}_{\{\mu\nu\}}$ and $J^{\bar D\Sigma_c^*}_{\{\mu\nu\}}$.

{\it It is important to note that}
although these pentaquark currents have definite parities ($3/2^-$ and $5/2^+$), they can couple to states of both positive and negative parities, by adding a $\gamma_5$ (see discussions in Refs.~\cite{Chung:1981cc,Jido:1996ia,Kondo:2005ur} and especially in Ref.~\cite{Ohtani:2012ps}):
\begin{eqnarray}
\langle 0 | J | B \rangle &=& f_{B} u(p) \, ,
\label{eq:gamma0}
\\ \langle 0 | J | B^\prime \rangle &=& f_{B^\prime} \gamma_5 u^\prime(p) \, ,
\label{eq:gamma5}
\end{eqnarray}
where $| B \rangle$ has the same parity as $J$, and $| B^\prime \rangle$ has the opposite parity. These equations also suggest that $J$ and $\gamma_5 J$ can couple to the same state,
and so the partners of these currents having opposite parities can also be used, such as $\gamma_5 \eta_{12\mu}^{c \bar c uud}$, but they just lead to the same sum rule results.

In this paper we shall use the non-$\gamma_5$ couplings, Eq.~(\ref{eq:gamma0}), and the couplings for $J^{\bar D^*\Sigma_c}_{\mu}$ and $J^{\bar D^*\Sigma^*_c}_{\{\mu\nu\}}$ are:
\begin{eqnarray}
\langle 0 |J^{\bar D^*\Sigma_c}_{\mu} | [\bar D^*\Sigma_c] \rangle &=& f_{\bar D^*\Sigma_c} u_\mu (p) \, ,
\\ \langle 0 |J^{\bar D^*\Sigma^*_c}_{\{\mu\nu\}} | [\bar D^*\Sigma_c^*] \rangle &=& f_{\bar D^*\Sigma^*_c} u_{\{\mu\nu\}} (p) \, .
\end{eqnarray}
The formulae are similar for $\eta_{12\mu}^{c \bar c uud}$, $\eta_{3\{\mu\nu\}}^{c \bar c uud}$, $J^{\bar D\Sigma_c^*}_{\mu}$, $J^{\bar D\Sigma_c^*}_{\{\mu\nu\}}$, and $J^{\bar D^*\Lambda_c}_{\{\mu\nu\}}$, which we shall not repeat.
Then the two-point correlation functions can be written as:
\begin{eqnarray}
\label{piDsS32}&& \Pi^{\bar D^*\Sigma_c}_{\mu \nu}\left(q^2\right)
\nonumber\\ \nonumber &&= i \int d^4x e^{iq\cdot x} \langle 0 | T\left[J^{\bar D^*\Sigma_c}_{\mu}(x) \bar J_{\nu}^{\bar D^*\Sigma_c} (0)\right] | 0 \rangle
\\ &&= \left(\frac{q_\mu q_\nu}{q^2}-g_{\mu\nu}\right) (q\!\!\!\slash + M_{[\bar D^*\Sigma_c]}) \Pi^{\bar D^*\Sigma_c}\left(q^2\right) + \cdots \, ,
\\ \label{piDsSs52} && \Pi^{\bar D^*\Sigma^*_c}_{\mu \nu \rho \sigma}\left(q^2\right)
\nonumber\\ \nonumber &&= i \int d^4x e^{iq\cdot x} \langle 0 | T\left[J^{\bar D^*\Sigma^*_c}_{\{\mu\nu\}}(x) \bar J_{\{\rho\sigma\}}^{\bar D^*\Sigma^*_c} (0)\right] | 0 \rangle
\\  &&= \left(g_{\mu\rho}g_{\nu\sigma} + g_{\mu\sigma} g_{\nu\rho} \right) (q\!\!\!\slash + M_{[\bar D^*\Sigma^*_c]}) \Pi^{\bar D^*\Sigma^*_c}\left(q^2\right) + \cdots \, ,
\end{eqnarray}
where the spin $1/2$ components are all contained in $\cdots$, such as $q_\mu q_\nu (q\!\!\!\slash + m) \Pi_{1/2}^{\bar D^*\Sigma_c}\left(q^2\right)$, etc..

One can also use the $\gamma_5$ couplings, Eq.~(\ref{eq:gamma5}). The resulting two-point correlation functions are similar to Eqs.~(\ref{piDsS32}) and (\ref{piDsSs52}), but with $(q\!\!\!\slash + M_X)$ replaced by $(- q\!\!\!\slash + M_X)$, where $X$ is either $[\bar D^*\Sigma_c]$ or $[\bar D^*\Sigma^*_c]$. This difference would tell us the parity of $X$. We note that the result does not change when using $\gamma_5 J^{\bar D^*\Sigma_c}_{\mu}$ and $\gamma_5 J^{\bar D^*\Sigma^*_c}_{\{\mu\nu\}}$ having opposite parities. Technically, in the following analyses we use the terms proportional to $\mathbf{1} \times g_{\mu\nu}$ and $\mathbf{1} \times g_{\mu\rho} g_{\nu\sigma}$ to evaluate the mass of $X$, which are then compared with those proportional to $q\!\!\!\slash \times g_{\mu\nu}$ and $q\!\!\!\slash \times g_{\mu\rho} g_{\nu\sigma}$ to determine its parity.

We follow Ref.~\cite{Chen:2015ata} and obtain $M_{[\bar D^*\Sigma_c]}$ and $M_{[\bar D^*\Sigma^*_c]}$ through:
%
\begin{eqnarray}
M^2_X(s_0, M_B) 
&=& {\int^{s_0}_{s_<} e^{-s/M_B^2} \rho^X(s) s ds \over \int^{s_0}_{s_<} e^{-s/M_B^2} \rho^X(s) ds} \, ,
\label{eq:mass}
\end{eqnarray}
%
where $\rho^X(s)$ is the QCD spectral density which we evaluate up to dimension eight, including the perturbative term, the quark condensate $\langle \bar q q \rangle$, the gluon condensate $\langle g_s^2 GG \rangle$, the quark-gluon mixed condensate $\langle g_s \bar q \sigma G q \rangle$, and their combinations $\langle \bar q q \rangle^2$ and
$\langle \bar q q \rangle\langle g_s \bar q \sigma G q \rangle$. The full expressions are lengthy and will not be shown here.
We use the values listed in Ref.~\cite{Chen:2015ata} for these condensates and the charm quark mass (see also Refs.~\cite{Yang:1993bp,Agashe:2014kda,Eidemuller:2000rc,Narison:2002pw,Gimenez:2005nt,Jamin:2002ev,Ioffe:2002be,Ovchinnikov:1988gk,colangelo}).

There are two free parameters in Eq.~(\ref{eq:mass}): the Borel mass $M_B$ and the threshold value $s_0$. We use two criteria to constrain the Borel mass $M_B$. One criterion is to require that the dimension eight term be less than 10\% to determine its lower limit $M_B^{min}$:
%
\begin{equation}
\label{eq_convergence}
\mbox{Convergence (CVG)} \equiv \left|\frac{ \Pi^X_{\langle \bar q q \rangle\langle g_s \bar q \sigma G q \rangle}(\infty, M_B) }{ \Pi^X(\infty, M_B) }\right| \leq 10\% \, ,
\end{equation}
%
and the other criterion is to require that the pole contribution (PC) be larger than 10\% to determine its upper limit $M_B^{max}$:
%
\begin{equation}
\label{eq_pole} \mbox{PC} \equiv \frac{ \Pi^X(s_0, M_B) }{ \Pi^X(\infty, M_B) } \geq 10\% \, .
\end{equation}
%
Altogether we obtain a Borel window $M_B^{min}<M_B<M_B^{max}$ for a fixed threshold value $s_0$. To determine  $s_0$, we require that both the $s_0$ dependence and the $M_B$ dependence of the mass prediction be the weakest.

We perform QCD sum rule analyses using $\eta_{12\mu}^{c \bar c uud}$ and $\eta_{3\{\mu\nu\}}^{c \bar c uud}$ of the $[\bar c_d c_d][\epsilon^{abc}q_a q_b q_c]$ configuration, but the results are not useful, because the spectral density $\rho^{[J/\psi N]}_{3/2}(s)$ obtained using $\eta_{12\mu}^{c \bar c uud}$ is too simple: it only contains the $q\!\!\!\slash \times g_{\mu\nu}$ part but no $\mathbf{1} \times g_{\mu\nu}$ part, and moreover, this $q\!\!\!\slash \times g_{\mu\nu}$ part only contains the perturbative term and $\langle g_s^2 GG \rangle$. There are also many terms missing in the spectral density $\rho^{[J/\psi N]}_{5/2}(s)$ obtained using $\eta_{3\{\mu\nu\}}^{c \bar c uud}$: its $q\!\!\!\slash \times g_{\mu\rho} g_{\nu\sigma}$ part only contains the perturbative term, $\langle g_s^2 GG \rangle$, $\langle \bar q q \rangle^2$ and $\langle \bar q q \rangle\langle g_s \bar q \sigma G q \rangle$, but its $\mathbf{1} \times g_{\mu\rho} g_{\nu\sigma}$ part only contains $\langle \bar q q \rangle$ and $\langle g_s \bar q \sigma G q \rangle$. This makes bad OPE convergence and leads to unreliable results.

\begin{figure}[hbt]
\begin{center}
\scalebox{0.33}{\includegraphics{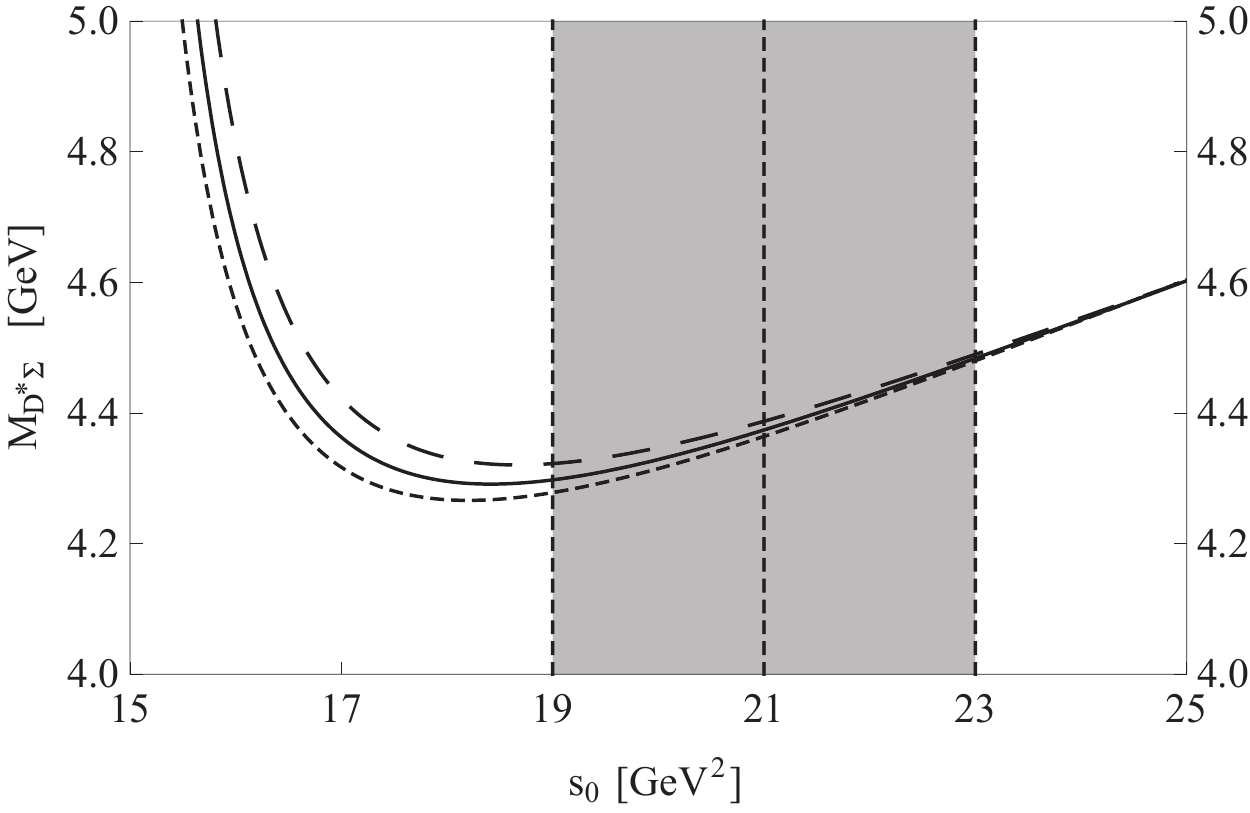}}
\scalebox{0.33}{\includegraphics{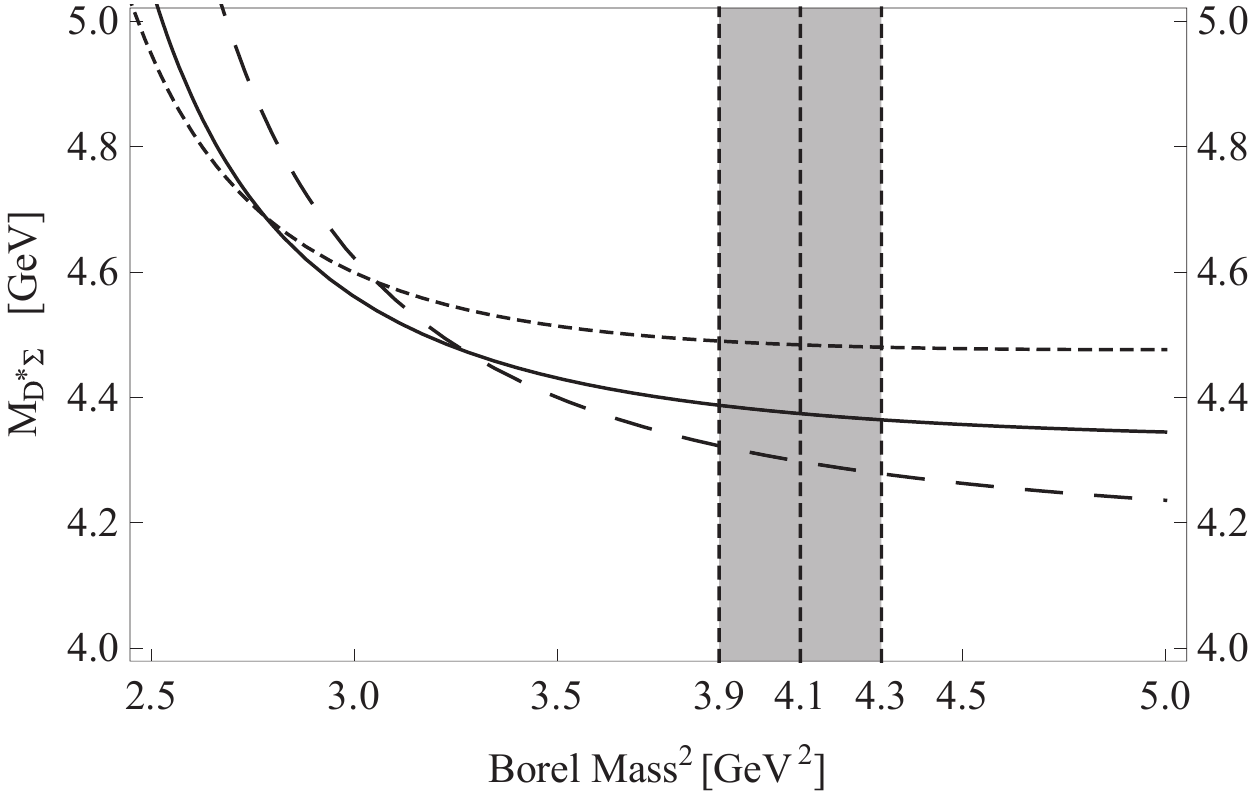}}
\caption{The variation of $M_{[\bar D^*\Sigma_c],3/2^-}$ with respect to the threshold value $s_0$ (left) and the Borel mass $M_B$ (right).
In the left figure, the long-dashed, solid and short-dashed curves are obtained by fixing $M_B^2 = 3.9$, $4.1$ and $4.3$ GeV$^2$, respectively.
In the right figure, the long-dashed, solid and short-dashed curves are obtained for $s_0 = 19$, $21$ and $23$ GeV$^2$, respectively.}
\label{fig:mass}
\end{center}
\end{figure}

We also perform QCD sum rule analyses using $J^{\bar D^*\Sigma_c}_{\mu}$, $J^{\bar D\Sigma^*_c}_{\mu}$, $J^{\bar D^*\Sigma_c^*}_{\{\mu\nu\}}$, $J^{\bar D\Sigma_c^*}_{\{\mu\nu\}}$, and $J^{\bar D^*\Lambda_c}_{\{\mu\nu\}}$ of the $[\bar c_d q_d][\epsilon^{abc}c_a q_b q_c]$ configuration. Here, we use $J^{\bar D^*\Sigma_c}_{\mu}[3/2^-]$, defined in Eq.~(\ref{DsS32}), as an example, whose sum rule has reasonable working regions. We calculate its spectral density, $\rho^{[\bar D^*\Sigma_c]}_{3/2}(s)$, and use its $\mathbf{1} \times g_{\mu\nu}$ part to evaluate the mass of $[\bar D^*\Sigma_c]$, denoted as $M_{[\bar D^*\Sigma_c]}$. We show its variation with respect to the threshold value $s_0$ in the left panel of Fig.~\ref{fig:mass}. We quickly notice that this dependence is the weakest around $s_0 \sim 18$ GeV$^2$, and the $M_B$ dependence is the weakest around $s_0 \sim 24$ GeV$^2$. Accordingly, we choose the region $19$ GeV$^2\leq s_0\leq 23$ GeV$^2$ as our working region. The corresponding Borel window is $3.9$ GeV$^2\leq M_B^2 \leq 4.3$ GeV$^2$ for $s_0 = 21$ GeV$^2$. We also show the variations of $M_{[\bar D^*\Sigma_c]}$ with respect to the Borel mass $M_B$ in the right panel of Fig.~\ref{fig:mass}, in a broader region $2.5$ GeV$^2\leq M_B^2 \leq 5.0$ GeV$^2$, while these curves are more stable inside the Borel window. We obtain the following numerical results:
\begin{eqnarray}
M_{[\bar D^*\Sigma_c]} = 4.37^{+0.19}_{-0.12} \mbox{ GeV} \, ,
\end{eqnarray}
where the central value corresponds to $M_B=4.1$ GeV$^2$ and $s_0 = 21$ GeV$^2$, and the uncertainty comes
from the Borel mass $M_B$, the threshold value $s_0$, the charm quark mass and the various condensates~\cite{Chen:2015ata}.
Finally, we find that the $q\!\!\!\slash \times g_{\mu\nu}$ part of the spectral density $\rho^{[\bar D^*\Sigma_c]}_{3/2}(s)$ is very
similar to the $\mathbf{1} \times g_{\mu\nu}$ part. This means that $[\bar D^*\Sigma_c]$ has the same parity as $J^{\bar D^*\Sigma_c}_{\mu}[3/2^-]$,
that is negative:
\begin{eqnarray}
M_{[\bar D^*\Sigma_c],3/2^-} = 4.37^{+0.19}_{-0.12} \mbox{ GeV} \, .
\label{Pc4380}
\end{eqnarray}
This value is consistent with the experimental mass of $P_c(4380)$~\cite{lhcb}, and supports it as a $[\bar D^*\Sigma_c]$ hidden-charm pentaquark with quantum numbers $J^P=3/2^-$.

The masses obtained using $J^{\bar D\Sigma_c^*}_{\{\mu\nu\}}[5/2^+]$ and $J^{\bar D^*\Lambda_c}_{\{\mu\nu\}}[5/2^+]$, defined in Eqs.~(\ref{DSs52}) and (\ref{DsL52}), depend much on the threshold value $s_0$ and so are not useful.
However, the following mixed current of $J^{\bar D\Sigma_c^*}_{\{\mu\nu\}}$ and $J^{\bar D^*\Lambda_c}_{\{\mu\nu\}}$ gives a reliable mass sum rule:
\begin{eqnarray}
J^{\bar D\Sigma_c^*\&\bar D^*\Lambda_c}_{\{\mu\nu\}} = \sin\theta \times J^{\bar D\Sigma_c^*}_{\{\mu\nu\}} + \cos\theta \times J^{\bar D^*\Lambda_c}_{\{\mu\nu\}} \, ,
\end{eqnarray}
when the mixing angle $\theta$ is fine-tuned to be $-51\pm5^\circ$, and the hadron mass can be extracted as
\begin{eqnarray}
 M_{[\bar D\Sigma_c^*\&\bar D^*\Lambda_c],{5/2^+}} = 4.47^{+0.20}_{-0.13} \mbox{ GeV} \, ,
 \label{Pc4450}
\end{eqnarray}
with $20$ GeV$^2$ $\leq s_0 \leq 24$ GeV$^2$ and $3.2$ GeV$^2$ $\leq M_B^2 \leq 3.5$ GeV$^2$. This value is consistent with the experimental mass of $P_c(4450)$~\cite{lhcb}, and supports it as an admixture of $[\bar D^* \Lambda_c]$ and $[\bar D \Sigma_c^*]$ with quantum numbers $J^P=5/2^+$. Accordingly to its internal structure described by $J^{\bar D\Sigma_c^*\&\bar D^*\Lambda_c}$, we suggest its main decay modes include $P$-wave $\bar D^* \Lambda_c$ and $\bar D \Sigma_c^*$ besides $J/\psi N$.

{\it The prediction of extra hidden-charm pentaquarks}.---
The tetraquark family can give us some information about the pentaquark family. To date, there are already six members in the family of the electrically charged states: $X(3900)^\pm$, $X(4020)^\pm$, $X(4050)^\pm$, $X(4250)^\pm$, $X(4430)^\pm$~\cite{Agashe:2014kda}, and $Z_c(4200)^+$~\cite{Chilikin:2014bkk}. They all contain at least four quarks, and can be described using the eight independent tetraquark currents with quantum numbers $I^GJ^{PC} = 1^+1^{+-}$, which represent internal structures of these states in the method of QCD sum rules (see Refs.~\cite{2011-Chen-p34010-34010,Chen:2015ata} and references therein). While, there are many independent pentaquark currents having quantum numbers $J=3/2$ and $J=5/2$, the more complicated internal structures of pentaquark suggesting that there may be more pentaquark states besides $P_c(4380)$ and $P_c(4450)$.

In this paper we use the pentaquark currents $J^{\bar D\Sigma_c^*}_{\mu}[3/2^-]$ and $J^{\bar D^*\Sigma_c^*}_{\{\mu\nu\}}[5/2^+]$, defined in Eqs.~(\ref{DsS32}) and (\ref{DsSs52}), to perform QCD sum rule analyses. Other currents of the same configuration ($[\bar c_d q_d][\epsilon^{abc}c_a q_b q_c]$) will be investigated in our future studies, where we shall do a systematical study in order to fully understand them. The mass obtained using $J^{\bar D\Sigma_c^*}_{\mu}[3/2^-]$ is
\begin{eqnarray}
M_{[\bar D\Sigma_c^*],{3/2}^-} &=& 4.45^{+0.17}_{-0.13} \mbox{ GeV} \, ,
\end{eqnarray}
and the mass obtained using $J^{\bar D^*\Sigma_c^*}_{\{\mu\nu\}}[5/2^+]$ is
\begin{eqnarray}
M_{[\bar D^*\Sigma_c^*],{5/2^+}} &=& 4.59^{+0.17}_{-0.12} \mbox{ GeV} \, .
\end{eqnarray}
Hence, we predict that there is the probability of a $[\bar D\Sigma_c^*]$ hidden-charm pentaquark having mass $4.45^{+0.17}_{-0.13}$ GeV and quantum numbers $J^P=3/2^-$ and a $[\bar D^*\Sigma_c^*]$ pentaquark having mass $4.59^{+0.17}_{-0.12}$ GeV and $J^P=5/2^+$. Accordingly to their internal structures described by $J^{\bar D\Sigma_c^*}_{\mu}$ and $J^{\bar D^*\Sigma_c^*}_{\{\mu\nu\}}$, we suggest that the former one $[\bar D\Sigma_c^*]$ mainly decay into $S$-wave $\bar D\Sigma_c^*$ and $J/\psi N$ and the latter one $[\bar D^*\Sigma_c^*]$ mainly decay into $P$-wave $\bar D^*\Sigma_c$ and $J/\psi N$.

If the hidden-charm pentaquarks exist in nature, there should be hidden-bottom pentaquarks with antibottom meson and bottom baryon components, which are as the partners of $P_c(4380)$ and $P_c(4450)$. Employing the previously obtained formalism, we further predict the masses of these possible hidden-bottom pentaquarks, i.e.,
\begin{eqnarray}
M_{[\bar B^*\Sigma_b],3/2^-} &=& 11.55^{+0.23}_{-0.14} \mbox{ GeV} \, ,
\label{Pb11550}
\\  M_{[\bar B\Sigma_b^*\&\bar B^*\Lambda_b],{5/2^+}} &=& 11.66^{+0.28}_{-0.27} \mbox{ GeV} \, .
\label{Pb11660}
\end{eqnarray}
The former one $[\bar B^*\Sigma_b]$ mainly will decay into $S$-wave $\Upsilon(1S)N/\Upsilon(2S)N$ and may decay into $\bar B^*\Sigma_b$, and the latter one $[\bar B\Sigma_b^*\&\bar B^*\Lambda_b]$ mainly decay into $P$-wave $\bar B\Sigma_b^*$, $\bar B^*\Lambda_b$, $\Upsilon(1S)N$, and $\Upsilon(2S)N$. These results provide valuable information for experimental exploration of these hidden-bottom pentaquarks.

{\it Conclusion}.---
In summary, the observation of $P_c(4380)$ and $P_c(4450)$ by LHCb \cite{lhcb} has opened a new window for studying hidden-charm exotic pentaquark states.

In this letter, we have performed a QCD sum rule investigation, by which $P_c(4380)$ and $P_c(4450)$ are identified as hidden-charm pentaquark states composed of an anti-charmed meson and a charmed baryon. We use $J_\mu^{\bar D^*\Sigma_c}$ to perform QCD sum rule analysis and the result shown in Eq.~\eqref{Pc4380} supports $P_c(4380)$ as a $[\bar D^*\Sigma_c]$ hidden-charm pentaquark with quantum numbers $J^P=3/2^-$. We use the mixed current $J^{\bar D\Sigma_c^*\&\bar D^*\Lambda_c}$ to perform QCD sum rule analysis, and the result shown in Eq.~\eqref{Pc4450} implies a possible mixed hidden-charm pentaquark structure of $P_c(4450)$, as an admixture of $[\bar D^* \Lambda_c]$ and $[\bar D \Sigma_c^*]$ with quantum numbers $J^P=5/2^+$, and its main decay modes include $P$-wave $\bar D^* \Lambda_c$ and $\bar D \Sigma_c^*$ besides $J/\psi N$.

Besides them, a) we use other two independent currents $J^{\bar D\Sigma_c^*}_{\mu}$ and $J^{\bar D^*\Sigma_c^*}_{\{\mu\nu\}}$ to perform QCD sum rule analyses, and predict there may be a $[\bar D\Sigma_c^*]$ hidden-charm pentaquark having mass $4.45^{+0.17}_{-0.13}$ GeV and quantum numbers $J^P=3/2^-$, and a $[\bar D^*\Sigma_c^*]$ hidden-charm pentaquark having mass $4.59^{+0.17}_{-0.12}$ GeV and $J^P=5/2^+$; b) we predict two hidden-bottom pentaquarks, as partners of $P_c(4380)$ and $P_c(4450)$. We also discuss their possible decay modes according to their internal structures described by pentaquark interpolating currents.

All these states/structures have a $[\bar c_d q_d][\epsilon^{abc}c_a q_b q_c]$ color configuration, could probe either a tightly-bound pentaquark structure or a molecular structure composed of an anti-charmed meson and a charmed baryon. We shall test more structures, such as the antiquark-diquark-diquark configuration, $\epsilon^{abc}[\bar c_a] [\epsilon^{bde}c_d q_e] [\epsilon^{cfg}q_f q_g]$, in our future studies.

In the near future, further experimental and theoretical study of hidden-charm/hidden-bottom (molecular) pentaquark will still be important, especially with the running of LHC at 13 TeV and forthcoming BelleII.

\section*{ACKNOWLEDGMENTS}

This project is supported by
the National Natural Science Foundation of China under Grants No. 11205011, No. 11475015, No. 11375024, No. 11222547, No. 11175073, and No. 11261130311,
the Ministry of Education of China (SRFDP under Grant No. 20120211110002 and the Fundamental Research Funds for the Central Universities),
and the Natural Sciences and Engineering Research Council of Canada (NSERC).

\end{document}